\documentstyle [12pt]{article}

\begin{document}

\hoffset = -1truecm
\voffset = -2truecm

\thispagestyle{empty}
\begin{flushright}
{\large \bf IC/93/328}\\
{\large \bf hep-th/9309148}\end{flushright}
\vskip 3truecm

\begin{center}

{ \bf DEFORMED HEISENBERG ALGEBRA AND FRACTIONAL SPIN FIELD IN 2+1
DIMENSIONS}\\
\vskip0.8cm
{ \bf
Mikhail S. Plyushchay\footnote{Permanent address:
IHEP, Protvino, Moscow Region, 142284 Russia,
e-mail: plushchay@mx.ihep.su}}\\[0.3cm]
{ International Centre for Theoretical Physics, I-34100 Trieste,
Italy}\\

\vskip2cm

                            {ABSTRACT}
\end{center}
With the help of the deformed Heisenberg algebra involving
Klein operator, we construct the minimal set of
linear differential equations for the (2+1)-dimensional
relativistic field with arbitrary fractional spin, whose value is
defined by the deformation parameter.

\vfill
\begin{center}
MIRAMARE--TRIESTE\\
September 1993
\end{center}
\newpage

The considerable interest to the (2+1)-dimensional
particles with fractional spin and
statistics (anyons) \cite{any} is conditioned nowadays by their applications
to the theory of planar physical phenomena: fractional quantum Hall effect
and high-$T_{c}$ superconductivity \cite{appl}.
Moreover, anyons attract a great attention due to their relationship
with different theoretical fields of research such as
conformal field theories, braid groups and deformed theories
(see, e.g., refs. \cite{fub}--\cite{ler}).

{}From the field-theoretical point of view, such particles can be
described in two, possibly related, ways. The first way consists in
organizing a statistical interaction of the scalar or fermionic field with
the Chern-Simons U(1) gauge field, that changes spin and statistics of
the matter field. In this approach, the gauge field can be
formally excluded from the theory due to their Lagrange equations (or
corresponding Hamiltonian constraints), resulting in the anyonic
permutation relations of the redefined matter field \cite{sem}.  But such an
exclusion was intensively criticized, and, till now it is not
clear whether the role of the Chern-Simons gauge field is reduced only to
the change of the spin and statistics of the matter field,
or there is some relic of the statistical gauge field in the theory
\cite{ger}.

Another, less developed way consists in attempting to describe
anyons within the group-theoretical approach analogously to the
case of integer and half-integer spin fields. The present paper is devoted
to further development of this approach,
whose program consists in constructing equations for
(2+1)-dimensional fractional spin field, subsequent identifying
corresponding field action and, finally, in realizing a
quantization of the theory to reveal a fractional statistics.
Within this approach, there are, in turn, two related possibilities: to
use many-valued representations of the $SO(2,1)$ group, or to
work with the infinite-dimensional unitary representations of
its universal covering group, $\overline{SO(2,1)}$ (or
$\overline{SL(2,R)}$, isomorphic to it) \cite{ply1}.  Up to now, only the
equations were proposed for the fractional spin field within the
former possibility \cite{ply1}-\cite{for}, whereas different variants
of the equations and field actions were constructed with the use
of the unitary infinite-dimensional representations of
$\overline{SL(2,R)}$ \cite{ply1}, \cite{ply3}-\cite{cor}.
At the same time, the problem of
quantizing the theory is still open here. The main difficulty
in quantization consists in the infinite component nature of the
fractional spin field which is used to describe in a covariant
way one-dimensional irreducible representations of the
(2+1)-dimensional quantum mechanical Poincar\'e group
$\overline{ISO(2,1)}$ specified by the values of mass and
arbitrary (fixed) spin. Due to this fact, an infinite set of the
corresponding Hamiltonian constraints must be present in the
theory to exclude an infinite number of the `auxiliary' field
degrees of freedom \cite{ply1}. This infinite set of constraints
should appropriately be taken into account.
But on the other hand, the infinite
component nature of the fractional spin field indicates on the
hidden nonlocal nature of the theory, and, therefore, can be
considered in favour of the existence of the anyonic
spin-statistics relation for the fractional spin fields within
the framework of the group-theoretical approach \cite{fro1,fro2}.

Relativistic field with arbitrary fractional spin $s=\varepsilon \alpha$,
$\alpha>0,$ $\varepsilon \pm1$,
can be described in this approach by the Klein-Gordon equation
\begin{equation}
(P^{2}+m^{2})\Psi=0
\label{kle}
\end{equation}
and (2+1)-dimensional analog of the Majorana equation \cite{maj}
\begin{equation}
(PJ-\varepsilon\alpha m)\Psi=0.
\label{maj}
\end{equation}
Here it is supposed that the field $\Psi=\Psi(x)$ transforms
according to the infinite-dimensional unitary irreducible representation
(UIR) of  the discrete series $D^{+}_{\alpha}$
or $D^{-}_{\alpha}$ of the group $\overline{SL(2,R)}$,
and $J^{\mu}$, $\mu=0,1,2,$ are the generators of $\overline{SL(2,R)}$ in
the corresponding representation, which satisfy the algebra
\begin{equation}
[J_{\mu},J_{\nu}]=-i\epsilon_{\mu\nu\lambda}J^{\lambda}
\label{alg}
\end{equation}
and the condition of the irreducibility,
$J_{\mu}J^{\mu}=-\alpha(\alpha-1)$ \cite{bar}.
Eqs. (\ref{kle}) and (\ref{maj}) fix simply the values
of the above-mentioned $\overline{ISO(2,1)}$  Casimir operators, which are
the operators of squared mass,  $M^{2}=-P^{2}$, and spin,
$S=PJ/m$, and moreover, they fix the energy sign: $sign\, P^{0}=\varepsilon
\varepsilon'$, where $\varepsilon'=+1$ and $-1$ for representations
$D^{+}_{\alpha}$ and $D^{-}_{\alpha}$, respectively  \cite{ply1,ply3}.
These equations are completely independent, unlike,
e.g., Dirac and Klein-Gordon equations for the case of
spinor field, and, therefore, are not very suitable
for constructing the action and quantum theory of the fractional
spin field.

In paper \cite{jac} Jackiw and Nair constructed linear differential
equations for the field with spin $s=\varepsilon(\alpha-1)$
using a direct product of the vector representation
and the UIRs of the discrete series $D^{+}_{\alpha}$ or $D^{-}_{\alpha}$,
whereas in ref. \cite{ply4} more simple set of
the Majorana-Dirac equations were proposed for the
field with spin $s=\varepsilon(\alpha-1/2)$ with the use of direct
product of the spinor and $D^{\pm}_{\alpha}$ representations.
In both cases the Klein-Gordon equation was a consequence of the basic
linear differential equations, but these equations comprised some
subsidiary conditions, which had to be introduced into the corresponding
field actions with Lagrange multipliers as some `external' conditions.

In paper \cite{stv} Volkov,  Sorokin and Tkach proposed
the spinor-like set of two linear differential equations
for the description of the spin-1/4 particles, whereas in ref.
\cite{sv} its supersymmetric extension was constructed to describe the
supermultiplet with the spin content $s=1/4$ and $3/4$.
It was done with the help of UIRs of the
discrete series $D^{+}_{\alpha}$ with $\alpha=1/4$ or $3/4$.
The generators $J^{\mu}$ in these two cases can be
represented in the form of operators bilinear in the bosonic oscillator
operators $a^{\pm}$.
It is this fact that allows to construct two
equations for one spin-$1/4$ field as a `square root' from the Majorana
and Klein-Gordon
equations, and to find a spinor representation of this
formulation. But, since for the case of arbitrary values $\alpha$
the above-mentioned realization for $J^{\mu}$ in terms of the bosonic
oscillator operators $a^{\pm}$ is impossible, the natural question arises:
is it possible to construct in any analogous way the equations for a field
with arbitrary fractional spin?

In recent paper \cite{cor} the vector set of three
linear differential equations
for a field with  arbitrary fractional spin was constructed. There
it was shown that any two of three equations
are independent and that the complete vector set of equations is necessary
only to have a covariant formulation. Besides, the proposed vector
set of dependent equations allowed to demonstrate that only the
representations of the discrete series $D^{\pm}_{\alpha}$ are appropriate
for the description of the fractional spin fields, whereas the possibility
to use the infinite dimensional UIRs
of the principal and additional continuous series of
$\overline{SL(2,R)}$ \cite{bar} is excluded.

Therefore, the problem of constructing the minimal covariant set  of
linear differential equations for the arbitrary fractional spin field is
still open, and the present paper is devoted exactly to the solution of
it.  For the purpose we use here
recently proposed realization of the generators $J^{\mu}$ in terms of the
operators $a^{\pm}$ forming the deformed Heisenbrg algebra involving the
Klein operator \cite{vas,bri},
and construct the generalization of the
spinor-like approach \cite{stv}
to the case of arbitrary fractional spin field.

So, let us consider the deformed Heisenberg algebra  \cite{bri,vas,mac}
\begin{equation}
[a^{-},a^{+}]=1+\nu Q,\quad Q^{2}=1,\quad
Qa^{\pm}+a^{\pm}Q=0,
\label{aa}
\end{equation}
with the self-conjugate Klein operator $Q$ and mutually conjugate
operators $a^{-}$ and $a^{+}$, and introduce the Fock-type vacuum,
\[
a^{-}|0>=0,\quad Q|0>=\kappa|0>,\quad
<0|0>=1,
\]
where $\kappa=+1$ or $-1$.
Then we get the action of the operator $a^{+}a^{-}$ on the
states $(a^{+})^{n}|0>$, $n=0,1,\ldots$,
\[
a^{+}a^{-}(a^{+})^{n}|0>=\left(n+\frac{1}{2}
\left(1+(-1)^{n+1}\right)\nu\kappa\right)(a^{+})^{n}|0>.
\]
Therefore, the  space of unitary representation of algebra (\ref{aa})
is given by the complete set of the normalized vectors
\begin{equation}
|e_{n}>=\left(\prod_{l=1}^{n}\left(l+\frac{1}{2}\left(1+(-1)^{l+1}
\right)\nu\kappa\right)\right)^{-1/2}(a^{+})^{n}|0>,
\label{en}
\end{equation}
\[
<e_{n}|e_{k}>=\delta_{nk},
\]
when $\nu\kappa>-1.$
Note that since the number operator is given here by
\[
N=a^{+}a^{-}-\frac{1}{2}\nu\kappa+\frac{1}{2}\nu Q,\quad
N|e_{n}>=n|e_{n}>,
\]
one can write down the Klein operator in the form
\[
Q=\kappa\cos\pi N,
\]
and obtain the transcendent equation for defining the operators $N$,
and hence $Q$,
in terms of the deformed creation and annihilation operators:
\[
\nu\kappa\cos\pi N=\nu\kappa+2(N-a^{+}a^{-}).
\]

With the help of the deformed Heisenberg algebra one can construct the
operators $J^{\mu}$ \cite{vas},
\begin{equation}
J_{0}=\frac{1}{4}(a^{+}a^{-}+a^{-}a^{+})=\frac{1}{2}a^{+}a^{-}
+\frac{1}{4}(1+\nu Q),\quad
J_{\pm}=J_{1}\pm iJ_{2}=\frac{1}{2}(a^{\pm})^{2},
\label{jmu}
\end{equation}
which form the algebra (\ref{alg}) of the generators of
$\overline{SL(2,R)}$ group.  The space of states (\ref{en})
is divided into two subspaces spanned by the states $|k>_{e}$
and $|k>_{o}$:
\[
|k>_{e}\equiv |e_{2k}>,\, \,
|k>_{o}\equiv |e_{2k+1}>,\, \, Q|k>_{e(o)}=+(-)\kappa|k>_{e(0)},\, \,
k=0,1,\ldots,
\]
invariant with respect to the action of operators (\ref{jmu}):
\begin{eqnarray}
J_{0}|k>_{e(o)}&=&(\alpha_{e(o)}+k)|k>_{e(o)},\quad k=0,1,\ldots,
\nonumber\\
J_{-}|0>_{e(o)}&=&0,\quad
J_{\pm}|k>_{e(o)}\propto|k\pm 1>_{e(o)},\, k=1,\ldots.
\label{inv}
\end{eqnarray}
The operator
\[
J_{\mu}J^{\mu}=-\frac{1}{4}(1+\nu Q)\left(\frac{1}{4}(1+\nu Q)-
1\right)
\]
takes the following values on these subspaces:
\begin{equation}
J_{\mu}J^{\mu}|k>_{e(o)}=-\alpha_{e(o)}(\alpha_{e(o)}-1)|k>_{e(o)},
\label{irreo}
\end{equation}
with
\begin{equation}
\alpha_{e}=\frac{1}{4}(1+\nu\kappa)>0,\qquad
\alpha_{o}=\alpha_{e}+\frac{1}{2}>\frac{1}{2}.
\label{aeo}
\end{equation}
Relations (\ref{inv}), (\ref{irreo})
mean that we have realized the UIRs of the
discrete series $D^{+}_{\alpha_{e}}$ and $D^{+}_{\alpha_{o}}$ of the group
$\overline{SL(2,R)}$ on the subspaces spanned by $|k>_{e}$
and $|k>_{o}$, respectively.
The UIRs of the discrete series $D^{-}_{\alpha_{e(o)}}$ can be obtained
from realization (\ref{jmu})
by means of the substitution
\begin{equation}
J_{0}\rightarrow -J_{0},\quad
J_{\pm}\rightarrow -J_{\mp},
\label{sub}
\end{equation}
and further, for the sake of simplicity, we shall consider only
the case of representations $D^{+}_{\alpha}$.

In order to construct the equations for the arbitrary fractional spin
field,
we introduce the $\gamma$-matrices in the Majorana representation,
\[
(\gamma^{0})_{\alpha}{}^{\beta}=-(\sigma^{2})_{\alpha}{}^{\beta},\quad
(\gamma^{1})_{\alpha}{}^{\beta}=i(\sigma^{1})_{\alpha}{}^{\beta},\quad
(\gamma^{2})_{\alpha}{}^{\beta}=i(\sigma^{3})_{\alpha}{}^{\beta},
\]
which satisfy the relation
$\gamma^{\mu}\gamma^{\nu}=-g^{\mu\nu}+i\epsilon^{\mu\nu\lambda}
\gamma_{\lambda}.$
Here $\sigma^{i}$, $i=1,2,3,$ are the Pauli matrices, and
raising and lowering the spinor
indices is realized  by the antisymmetric tensor
$\epsilon_{\alpha\beta}$, $\epsilon_{12}=\epsilon^{12}=1$:
$f_{\alpha}=f^{\beta}\epsilon_{\beta\alpha},$
$f^{\alpha}=\epsilon^{\alpha\beta}f_{\beta}.$

Let us consider now the spinor operator
\begin{eqnarray}
L_{\alpha}=
\left(\begin{array}{ccc}
q\\
p
\end{array}\right),
\label{lab}
\end{eqnarray}
constructed from the  operators $q$ and $p$, which
are defined by the operators $a^{\pm}$:
\[
a^{\pm}=\frac{q\mp ip}{\sqrt{2}},
\]
and in correspondence with eq. (\ref{aa}) satisfy the algebra
\[
[q,p]=i(1+\nu Q),\quad Qq+qQ=Qp+pQ=0.
\]
Taking into account equalities (\ref{jmu}), we find that the operators
\begin{displaymath}
J_{\alpha\beta}\equiv i(J_{\mu}\gamma^{\mu})_{\alpha\beta}=
\left(\begin{array}{ccc}
J_{0}+J_{1} & -J_{2}\\
-J_{2} & J_{0}-J_{1}
\end{array}\right)
\end{displaymath}
can be represented in the following form:
\[
J_{\alpha\beta}\equiv \frac{1}{4}(L_{\alpha}L_{\beta}+
L_{\beta}L_{\alpha}).
\]
This equality means that operator (\ref{lab}) is the `square root'
operator of the $\overline{SL(2,R)}$ generators (\ref{jmu}).

With the help of operator (\ref{lab}),
construct the spinor linear differential operator
\begin{equation}
D_{\alpha}=L^{\beta}\left((P\gamma)_{\beta\alpha}-\varepsilon
m\epsilon_{\beta\alpha}\right),
\label{da}
\end{equation}
where $P_{\mu}=-i\partial_{\mu}$, $\varepsilon=\pm1$,
and consider the spinor pair of equations
\begin{equation}
D_{\alpha}\Psi=0.
\label{baseq}
\end{equation}
88We suppose here that the field $\Psi=\Psi_{e}(x)$ or
$\Psi=\Psi_{o}(x)$, is decomposable into the series
\begin{equation}
\Psi_{e(o)}(x)=
\sum_{k=0}^{\infty}\psi^{k}_{e(o)}(x)|k>_{e(o)}.
\label{}
\end{equation}
Operator (\ref{da}) satisfies the relation
\[
D^{\alpha}D_{\alpha}=L^{\alpha}L_{\alpha}(P^{2}+m^{2}),
\]
which together with the relation
$L^{\alpha}L_{\alpha}=-i(1+\nu Q)\neq 0$ means that
the Klein-Gordon equation (\ref{kle}) is the
consequence of eqs. (\ref{baseq}). Moreover, due to the relation
\[
L^{\alpha}D_{\alpha}=-4i\left(PJ-\varepsilon m\frac{1}{4}(1+\nu Q)\right),
\]
we conclude that the fields $\Psi_{e}$ and $\Psi_{o}$
satisfying eqs. (\ref{baseq}), obey also the corresponding equations
\begin{equation}
(PJ-\varepsilon\alpha_{e}m)\Psi_{e}=0,
\label{maje}
\end{equation}
\begin{equation}
\left(PJ-\varepsilon
\left(\alpha_{o}-\frac{1}{2}(1+\nu\kappa)\right)m\right)\Psi_{o}=0.
\label{majo}
\end{equation}
Eq. (\ref{maje}) is nothing else as the Majorana equation (\ref{maj}).
Taking into account eq. (\ref{kle}), and
passing over to the rest frame ${\bf P}={\bf 0}$ in the momentum
representation, we find with the help of eq. (\ref{jmu}) that eq.
(\ref{maje}) has the
solution of the form $\Psi_{e}\propto \delta(P^{0}-\varepsilon m)
\delta({\bf P})
\psi^{0}_{e}|0>_{e}$, whereas eq. (\ref{majo}) has no nontrivial solution.
Therefore, the pair of equations (\ref{baseq})
has nontrivial solutions only in
the case $\Psi=\Psi_{e}$, i.e.  when $Q\Psi=\kappa\Psi$,
describing the field with spin $s=\varepsilon\alpha_{e}$ and mass $m$.
At $\nu=0$, i.e. when $[a^{-},a^{+}]=1$,
eqs. (\ref{baseq}) in correspondence with eqs. (\ref{aeo}), (\ref{maje}),
turn into Volkov-Sorokin-Tkach equations for a
field $\Psi_{e}$ with spin $s=1/4\cdot \varepsilon$ \cite{stv}
being analogous to (3+1)-dimensional Dirac
positive-energy relativistic wave equations \cite{dir}.

Thus, we have constructed the spinor-like system of
linear differential equations (\ref{baseq}) being
the independent equations for the field
$\Psi=\Psi_{e}$ with arbitrary fractional spin
defined by  the deformation parameter:
$s=\varepsilon\cdot\frac{1}{4}(1+\nu\kappa)\neq 0$.
Eqs. (\ref{baseq}) are in fact the `square root' of the Klein-Gordon and
Majorana equations and they can be used as a
starting point to construct the quantum theory
of the anyons within the group-theoretical approach.
In future publications we hope to investigate the problems of
constructing the field action leading to equations (\ref{baseq}) and
quantization  of the theory, and, besides,
we will consider the supersymmetric extension of
the proposed formulation for the arbitrary fractional spin field.

\vskip \baselineskip

{\bf Acknowledgements.}

The author thanks Prof. Abdus Salam, the International Atomic Agency and
${\rm UNESCO}$ for hospitality at
the International Centre for Theoretical Physics, Trieste, where the
present work was completed.

He is also grateful to A.T. Filippov, P.A. Marchetti, J. Myrheim and
M.V. Vasiliev for useful discussions,
and to M. Tonin for discussions and hospitality
at Dipartimento di Fisica, Universit\'a di Padova.

\end{document}